# Rapid Assessment of Stable Crystal Structures in Single Phase High Entropy Alloys Via Graph Neural Network Based Surrogate Modelling

Nicholas Beaver [a], Aniruddha Dive [a], Marina Wong [a], Keita Shimanuki [a], Ananya Patil [a], Anthony Ferrell [a], Mohsen B. Kivy [a]

a: Materials Engineering Department, California Polytechnic State University, 1 Grand Ave, San Luis Obispo, CA 93407, USA

**Abstract**

In an effort to develop a rapid, reliable, and cost-effective method for predicting the structure of single-phase high entropy alloys, a Graph Neural Network (ALIGNN-FF) based approach was introduced. This method was successfully tested on 134 different high entropy alloys, and the results were analyzed and compared with density functional theory and valence electron concentration calculations. Additionally, the effects of various factors, including lattice parameters and the number of supercells with unique atomic configurations, on the prediction accuracy were investigated. The ALIGNN-FF based approach was subsequently used to predict the structure of a novel cobalt-free 3d high entropy alloy, and the result was experimentally verified.

## 1. Introduction

For centuries, conventional alloy design had been based on one principal element in the alloy systems with addition of some other minor alloying elements to enhance their property/performances. Even the development of metal-matrix composites in 1970's relied mainly on conventional alloys as the matrix [1, 2]. In the early 21$^{st}$ century, high-entropy alloys (HEAs) as a novel concept of practical alloying were introduced by Cantor *et al.* and Yeh *et al.* independently [3, 4]. HEAs are a class of multi-principal alloys with four or more constituent elements, and equi-atomic or near equi-atomic elemental ratios. Due to their unique composition, HEAs are also referred to as complex, concentrated alloys (CCAs) or multi-principal element (MPE) alloys [5, 6]. Despite having multiple principal elements, high configurational and mixing entropy in HEAs result in thermodynamic stability of simple solid solutions (SS) in their microstructures. The extensive review study done by D. B. Miracle and O. N. Senkov showed that the majority of studied HEAs form FCC, BCC, B2 or a combination of them. Finding other phases such as HCP, σ-phase, or intermetallic in HEAs are not as common [6]. The vast compositional potential and the simple microstructures have enabled researchers to develop HEAs with promising and exceptional properties such as high strength, high corrosion resistance, high wear resistance, high-temperature mechanical performances, and recently high hydrogen storage capacity [7-9]. Despite demonstrating significant potential, complexity and design intricacy of HEAs have limited the exploration of these materials. Traditionally, the design and fabrication of HEAs relied on extensive trial and error, employing empirical guidelines that yield mixed success and accidental discoveries [10-12]. To address the challenges in the experimental design of HEAs such as minimizing the number of laborious trial-and-errors or reducing the inherent costs of empirical investigations, various computational modeling and simulation tools at different length or time scales have been developed or employed to predict the microstructure and/or properties of HEAs [13]. Although these computational tools show promising outcomes in studying the HEAs, their practicality or accessibility remains limited for many industries or universities. For instance, first principles approach is limited to only a few hundred atoms and can be computationally expensive; molecular dynamics typically uses open-source software but only a very few interatomic potentials developed for HEAs are available; and reliable HEA CALPHAD databases can be very costly to acquire. The most popular amongst these methods are the ones that utilize valence electron concentration (VEC) to predict the crystal structure of single-phase HEAs [10, 14-17]. This approach works very well for HEAs where the VEC values fall within the range corresponding to FCC and BCC respectively. However, when the VEC values lie in between the ranges, the algorithm fails, and the prediction accuracy drops significantly. Therefore, it seems to be essential to investigate new theoretical tools to study HEAs more quickly, reliably, and cost-effectively. To address the above challenges, in the present study we aim to develop a simple and fast approach to rapidly predict the stable structure of HEAs through ML-based structure predictions.

Over the past decade, there has been a remarkable surge in the utilizing machine learning (ML) – based approaches towards discovery of novel alloys. The progress of ML-based methodologies includes predicting stable phases (microstructure) and properties, accelerating simulations, and extracting underlying physical properties from the complex chemical structure of HEAs [16, 18-22]. Properties of these HEAs heavily rely on the phase compositions i.e., whether the HEA is a single-phase solid solution (SS) or multiphase. Existing literature reports multitude of studies predicting a wide range of stable phases in HEAs and report the impact of change in phases on respective mechanical and physical properties [23-29]. However, predicting stable phases within HEAs can be extremely challenging firstly due to wide range of empirical parameters and secondly due to their distinct sensitivity towards predicting the stable phases [30-32]. Consequently, the accurate identification and characterization of mixtures involving distinct phases have not been extensively addressed. Even within the design domain of single-phase SS, existing methodologies that can predict the stable crystal structure of single-phase SS HEAs are limited and lack accuracy. Furthermore, most of these studies relied on small datasets containing fewer than 1000 experimental records with similar features.

Employing machine learning approaches using composition-based embedding have been reported in the literature for discovering and optimizing HEAs [33, 34]. R. Bobbili and B. Ramakrishna have reported that decision trees can be used to rediscover HEA phase formation rules through atomic size difference and such decision trees can be used to accurately predict the phase formation of the HEAs in each dataset [35]. While methods such as these are useful for extracting patterns from data, they are inherently interpolative methods that rely on existing data to be useful. For instance, Cantor estimated that as many as $10^{100}$ unique HEA compositions can be designed [36], which means that much more data on HEAs should be needed to learn general trends about phase formation using composition-based embeddings. Ideally, density functional theory (DFT) could be used to generate data at the atomic level that can be used in useful downstream tasks that rely on data like the ML models discussed above. However, the high cost of DFT calculations create a need for cheaper models that can support atomic exploration and data creation. The frontrunner for surrogate atomic modelling is the Graph Neural Network (GNN). GNNs can encode the geometry and many body interactions of atomic structures implicitly, leading to improved predictions at the structural level [37-39]. Specifically, training a GNN on the energy of atomic structures as calculated from DFT can be used to make universal interatomic potentials, which was independently developed by the teams of Chi Chen, Shyue Ping Ong [40] and Choudhary *et. al* [41]. Importantly, Chen and Ong showed that these universal potentials can be used as a DFT surrogate to filter tens of millions of materials based on hull energy as an upstream task for more expensive DFT calculations. This type of task is essential for investigating the vast permutation space of HEAs. Choudhary *et al.* assert that their universal potential should be valid for materials systems including HEAs, therefore we deployed their Atomistic Line Graph Neural Network-based Force-Field (ALIGNN-FF) ML potentials in our present work and developed a quick and reliable approach to determine the stable crystal structure of HEAs through just the chemical composition of the HEAs.

Our proposed approach scans through complex potential energy surface (PES) for any given composition of HEA and utilizes ALIGNN-FF to determine the stable crystal structure. Since the majority of the HEAs are reported to form either FCC or BCC, only these structures were considered in this work. Calculating the ground-state cubic crystal structure of a given HEA composition requires finding a representative atomic permutation for the distinct supercell structure of that composition. To avoid computationally expensive algorithm such as Special Quasirandom Structure (SQS) generators, genetic algorithms, generative algorithms, and simulated annealing in combination with DFT, our method used Monte Carlo sampling to generate several possible combinations to find the best atomic arrangement and calculate the energies of FCC and BCC of HEAs. The accuracy of the method was tested in comparison with experimental data in the literature as well as the VEC predictions. It was then used to find a brand-new single-phase HEA with desired composition and structure, and the result was verified with experiment.

## 2. Materials and Methods

*Figure 1* provides a brief overview of the proposed approach in this work to predict the crystal structure for any given HEA composition. It begins by generating a 3×3×3 supercells utilizing the experimentally reported lattice parameters (3.54 Å for FCC, 3 Å for BCC), then these supercells are randomly populated with respective atoms in each composition of alloy. A Monte Carlo approach was employed to generate 500 combinations of distinct supercells for each FCC and BCC structures by randomly interchanging the positions of atoms in each configuration, and it was ensured that no two configurations were exactly similar. A total of 132 single-phase HEAs were selected from the literature where their structures were reported experimentally (the list of these HEAs is included in the Appendix). To calculate the normalized energies (eV/atom) and determine the lowest energy structures, ALIGNN-FF was then applied to each generated supercell with unique atomic configurations and was repeated 9 times to eliminate any calculation error. This resulted in 9000 iterations per HEA.

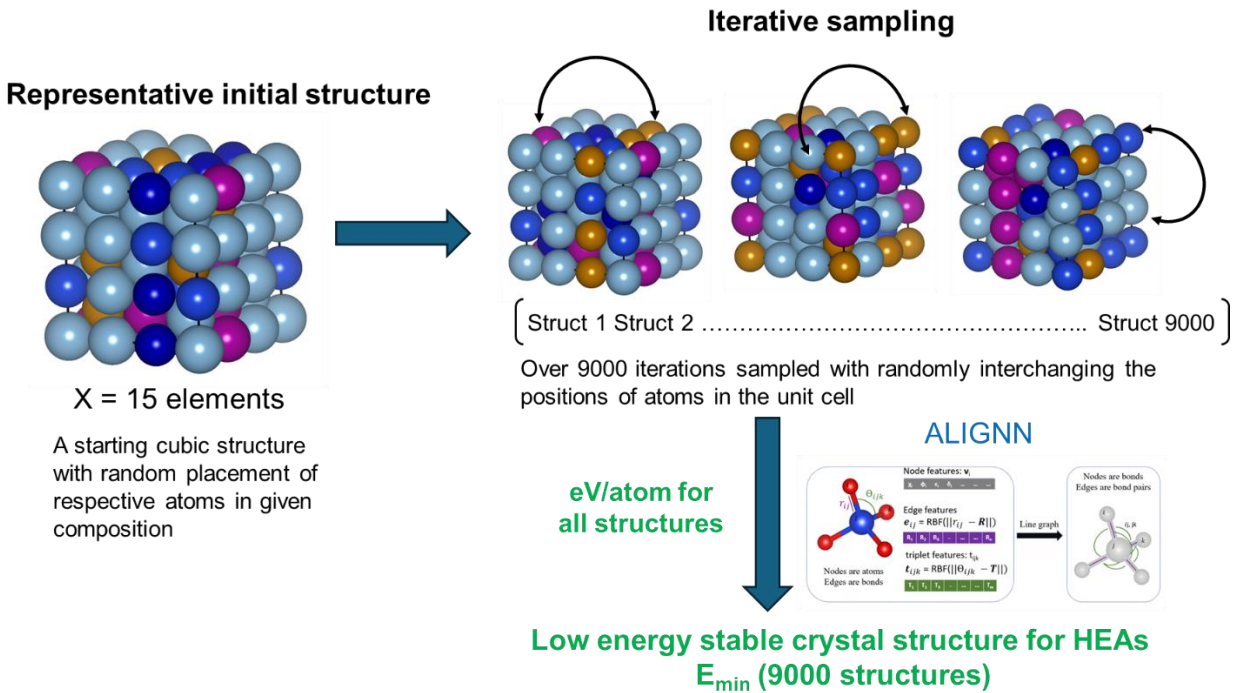

*Figure 1.* The flowchart summarizing our approach for rapidly predicting the stable crystal structure of HEAs. The first step involves generating a supercell with random atomic configuration for a selected composition of HEA. A Monte Carlo approach is used to sample 1000 configurations in the design space by randomly interchanging the positions of atoms in the supercell. Lastly, ALIGNN-FF was utilized to predict the lowest energy stable crystal structure for any given composition of HEAs.

*Figure 2* shows the sample BCC and FCC structures that were generated for Al$_2$FeCuNiMn, using the method described above. For DFT calculations, the generalized gradient approximation (GGA) with the projector-augmented wave (PAW) pseudopotentials implemented in the Vienna Ab-Initio Simulation Package (VASP) were used. The selected plane-wave cutoff was set at 600 eV with a gamma-point k-sampling. The DFT energies and atomic forces were converged to within 0.01 meV/atom and $10^{-2}$ eV·Å$^{-1}$, respectively.

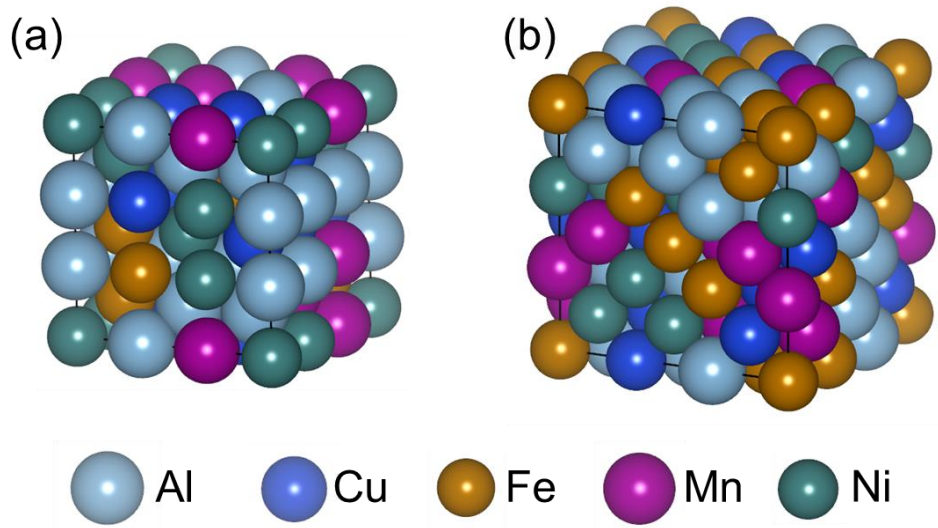

*Figure 2*. The representative starting configuration for Al$_2$FeCuNiMn alloy as generated by our approach in (a) BCC and (b) FCC supercell.

For the experiments, alloying elements with 99.95% purity or higher were weighed, stirred with ethanol for cleaning and dehydration, pre-mixed, and cast using an MRF SA-200 vacuum arc melting furnace. All casting was done under argon gas with atmospheric pressure, and the samples were remelted multiple times to ensure their homogeneity. The cast samples were sectioned, ground, polished, and thoroughly cleaned in an ultrasonic bath. X-ray Fluorescence (XRF) was done using Rigaku NEX DE VS instrument. Rigaku Miniflex X-ray Diffraction (XRD) with Cu Kα source and a Rigaku Haskris air-cooled chiller was used to identify the phases. Samples were mounted on a bulk sample holder; detection angle range was set to 20°-90° with a step increment of 0.05° and speed of 0.7° per minute, and the results were then analyzed using Rietveld method. FEI Quanta 200 Scanning Electron Microscope (SEM) with 20-25 kV and spot size of 4 was used for microstructure imaging. Each experimental characterization was repeated for multiple samples to ensure the accuracy of observations.

## 3. Results and Discussions

As initial verification of the ALIGNN-FF in calculating the energies of HEA structures, it was applied to the selected 132 single-phase HEAs. For each alloy, three FCC and three BCC supercells were generated with unique atomic positions, then the energy of each supercell was calculated using ALIGNN-FF and DFT independently. As shown in *Figure 3*, although the resulting 792 energies predicted by ALIGNN-FF were consistently higher compared to those obtained through DFT calculations (which was expected since the ALIGNN-FF potentials were not trained on the above DFT data); the quantitative trends in the energies between different configurations were still captured accurately by ALIGNN-FF. Moreover, the computational cost for performing ALIGNN-FF energy calculations for these configurations was calculated to be ~100× lower than that for DFT calculations. As a result, ALIGNN-FF was positively chosen to continue with the subsequent computations.

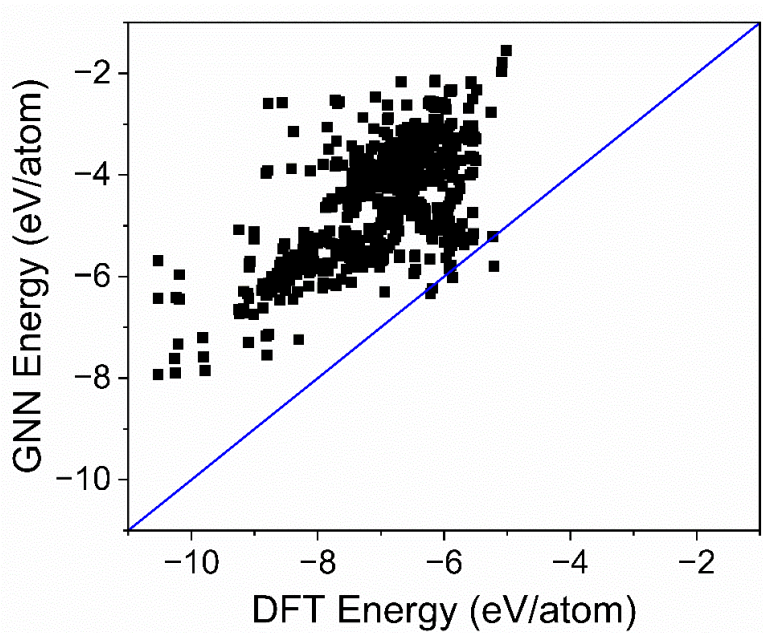

*Figure 3*. Calculated configurational energy for 792 distinct structures using DFT and ALIGNN-FF

To further ensure the veracity of the ALIGNN-FF method predictions, for each of the 132 selected single-phase HEA, 500 supercells with unique atomic positions for each FCC and BCC were generated and then the energy calculation was performed on each supercell for 9 times. The resulting 9000 energy values were then compared, and the lowest energy was used to determine the stable structure of each HEA. The structures of the selected 132 single-phase HEAs were also predicted through the VEC descriptor (LSS1). To determine how accurate and successful each method was, the prediction results of both methods were then compared to the experimental data in the literature. *Figure 4* shows the prediction accuracy of the ALIGNN-FF and the LSS1 methods determined by experimental data. In this figure, the blue datapoints represent alloys where the GNN and/or LSS1 prediction matched the experimental data of the alloy ("success"), and red datapoints represent compositions where they did not match ('fail"). The top portion of the figure is dedicated to the LSS1 predictions, while the bottom portion is for ALIGNN-FF. The 132 HEAs are on the horizontal axis; the alloys on the right of the dashed line are those which ALIGNN-FF method successfully predicted, while the ones on the left of the dashed line are the alloys were the prediction of ALIGNN-FF was different from the experiment. All the calculated energy values as well as the predictions are included in the appendix.

As shown in *Figure 4*, using ALIGNN-FF to predict the ground state structure of single-phase HEAs resulted in a higher accuracy (82%) compared to LSS1 method (79%). Due to the discontinuity between the crystal structure determination criteria values, VEC descriptor LSS1 methods predicted 28 HEAs to have a binary structure as compared to the single-phase structures determined by experiments. On the other hand, the ALIGNN-FF showed better accuracy in predicting the stable crystal structures for these HEAs since there is no discontinuity in the determination criteria. The majority of the alloys for which LSS1 accurately predicted their structures, ALIGNN-FF approach was also successful.

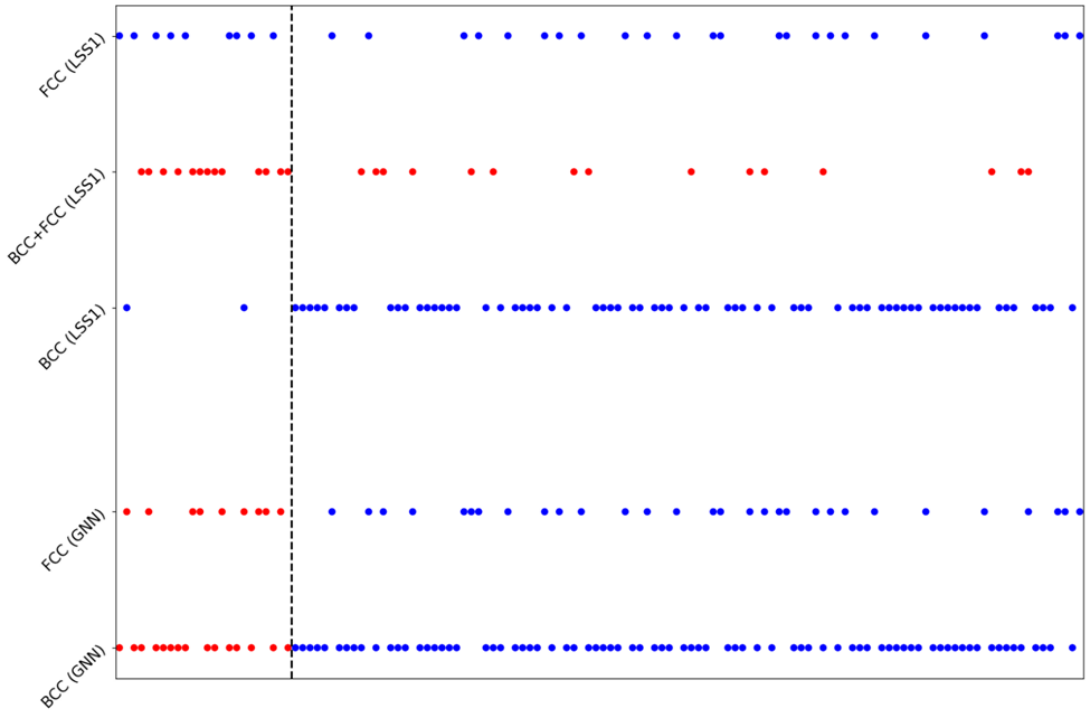

*Figure 4*. The comparison of stable crystal structures for 132 HEAs predicted by ALIGNN-FF (bottom two rows) and LSS1 (top three rows). Blue datapoints represent the alloys wherein the predicted structures matched experimental data. Red datapoints represent the alloys wherein the predicted structure did not match the experimentally reported structures.

To better understand the source of error (fails) in ALIGNN-FF predictions, the calculated energies of all FCC and BCC supercells for each HEA were carefully analyzed. The GNN criterion is based on identifying the configurations with lowest possible energy amongst the different sampled configurations. Therefore, the mean energy for the FCC and BCC sampled supercells for each alloy were calculated and plotted as *Figure 5*. In this figure, the calculated mean energy ($E_m$) for 132 HEAs in both FCC and BCC configurations have been plotted against the 132 selected HEAs (horizontal axis). The red datapoints on the left side of the dashed line represent the incorrect predictions and the blue datapoints show correct predictions. It can be clearly observed that amongst all the correct predictions, mean energies for FCC and BCC configurations are well separated or have significantly distinct values ($E_m > 0.5$ eV/atom). However, for the incorrect predictions, the mean energy between FCC and BCC configurations is almost equivalent ($E_m < 0.002$ eV/atom) for majority of the alloys. Such a small difference in the mean energies of the BCC and FCC configurations could possibly suggest a lack of sampling for these configurations, resulting in inaccurate prediction of their structures. As stated earlier, ALIGNN-FF potentials have not been trained on the HEA DFT energies and therefore such narrow energy differences between different structures could result in discrepancies in the predictions.

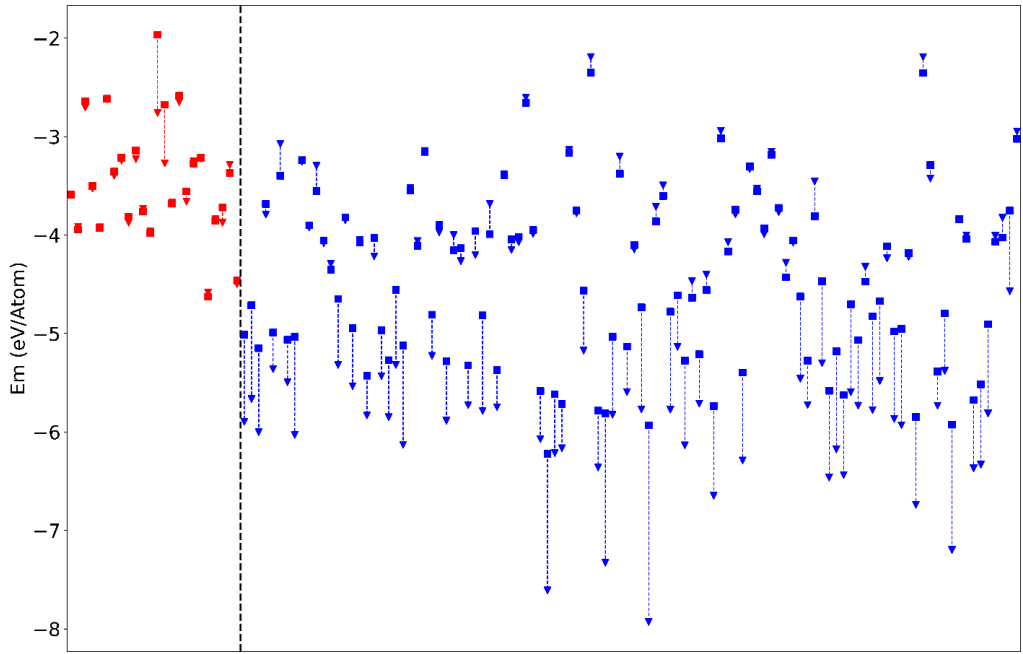

*Figure 5.* Calculated mean energies ($E_m$) for selected 132 HEAs. FCC and BCC supercells are represented as square and triangular datapoints respectively. The dashed lines show the difference between the mean energies for FCC and BCC configurations that were sampled. Blue datapoints represent the alloys wherein the predicted structures matched experimental data. Red datapoints represent the alloys wherein the predicted structure did not match the experimentally reported structures.

The results presented in *Figures 4* and *5* were based on sampling a total of 4500 (500 supercells × 9 iterations for each supercell) configurations for each FCC and BCC structures, given the composition of alloys. To understand the impact of change in the sampling number on the prediction accuracy, the above methodology was repeated for 1, 5, 10, 25, 50, 100 and 250 number of supercells with unique atomic positions per iteration, for a total of 9 iterations, as described in *Figure 1*. The accuracy/success of each of these sampling frequencies in predicting the structure of the selected 132 HEAs have been shown in *Figure 6*. As can be seen, the accuracy of the prediction observes a gradual increase with increasing the number of sampling frequencies. This is because larger sampling size enables the search through a more realistic potential energy surface and determine a stable crystal structure for any given alloy. Moreover, *Figure 6* revealed that the increase in accuracy plateaued out beyond 100 sampling iterations. Thus, increasing sampling iterations above 100 configurations did not show any significant advantage but adversely increased the computational cost. As a result, the optimal number of sampling iterations to achieve the best results for rapid and accurate prediction of stable structures is determined to be 100.

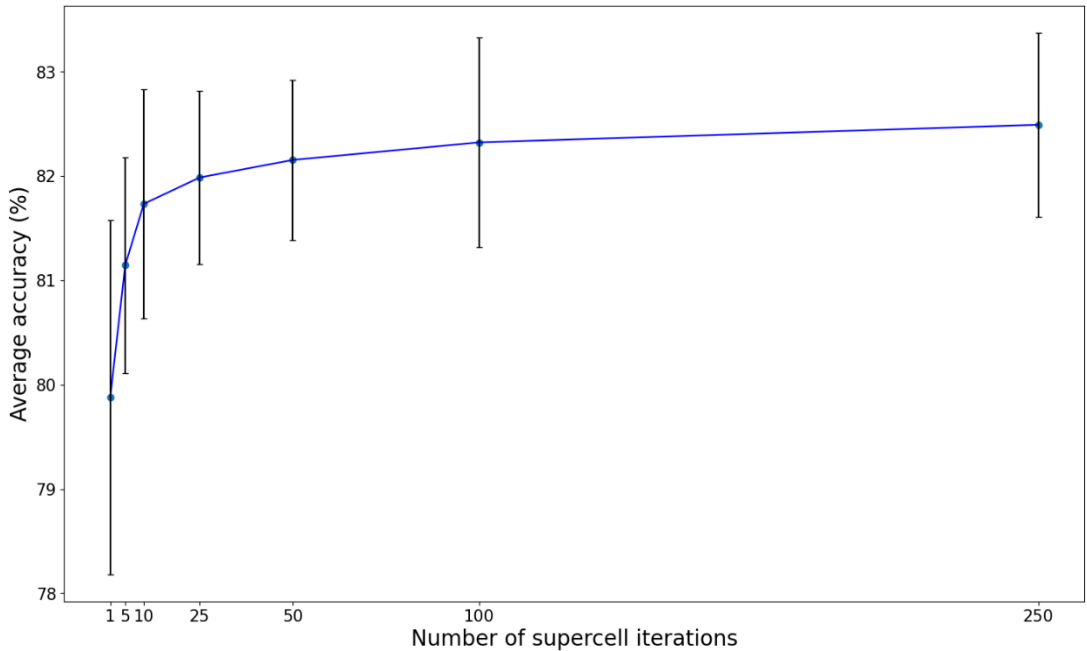

*Figure 6.* Calculated average accuracy for predicting the structures of selected 132 HEAs as a function of sampling iterations. The error bars represent the accuracy for 9 iteration runs for each supercell.

After examining the impact of sampling iterations on the accuracy of prediction and identifying an optimum number of iterations, the effect of lattice parameters (supercell sizes) was also investigated. Restricting the configuration space to a fixed initial supercell volume does not capture the realistic potential energy surface and might introduce some bias towards stable structure prediction. To consider the impact of initial supercell volume on the accuracy of predictions, different lattice parameters for both FCC and BCC supercells were considered and our ALIGNN-FF based approach was applied. The heatmap in *Figure 7* shows the impact of changing the lattice parameters for FCC and BCC supercells on the accuracy of predicting the stable structures of the selected 132 HEAs. As indicated in *Figure 7*, the lattice parameter of the initial supercell showed an impact on the accuracy of the prediction. Although the initial lattice parameters of 3.0 Å for BCC supercell and 3.54 Å for FCC supercells did not yield the best accuracy, the results were only 1% different compared to the highest achieved accuracy. All other lattice parameters combinations, outside the ones shown in *Figure 7* resulted in lower accuracy than 73%.

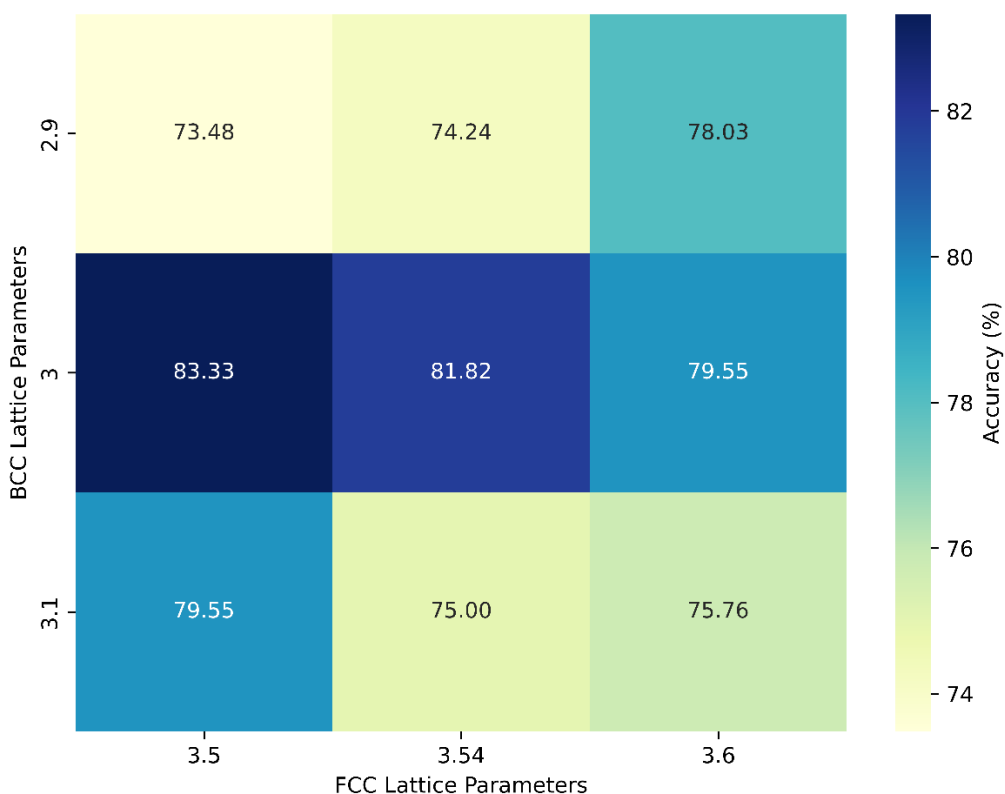

*Figure 7.* The average accuracy of prediction of stable structures of selected 132 HEAs as a function of lattice parameters for FCC and BCC supercells.

To showcase the applicability of ALIGNN-FF and LSS1 in searching for unknown and unstudied single-phase HEAs, both methods were tested to find an alloy with single-phase BCC structure. Upon a thorough literature review, $Al_2FeCuNiMn$ quinary HEA was selected as a candidate due to two main reasons: (a) single-phase BCC HEA design has been primarily based on refractory elements [6, 42]. However, these RHEAs can have limited applicability due to their high densities, brittle fracture behavior, and high manufacturing cost due to the elemental cost and high melting temperatures [10, 43, 44]. Therefore, developing 3d transition BCC HEAs can extend the pertinency of these alloys for a wider range of applications. (b) majority of studied 3d HEAs in the literature have been based on the original Yeh alloy (Al$_x$(CrFeCoNiCu)), containing cobalt as one of the main alloying elements [45, 46]. Although cobalt has been shown to have positive effects on increasing the strength of HEAs, it has shown to be an FCC stabilizer, and not advantageous when developing a BCC HEA [47-49]. Moreover, cobalt is widely known for its toxic potential that can lead to various health problems, such as allergic dermatitis, rhinitis, and asthma [50, 51]. Prior to experimentally casting the Al$_2$FeCuNiMn alloy, both GNN and LSS1 methods were applied to predict its microstructure and the results have been shown in *Table 1*.

*Table 1.* Predicted stable crystal structure of Al$_2$FeCuNiMn using GNN and LSS1 models is shown.

| Method | ALIGNN-FF | LSS1 |
| --- | --- | --- |
| Predicted Microstructure | BCC | FCC+BCC |

As can been in *Table 1*, the GNN method predicted a single-phase BCC for Al$_2$FeCuNiMn, while LSS1 predicted a binary FCC+BCC phase. This is because the VEC=7 of Al$_2$FeCuNiMn which is in the mixed FCC+BCC region [17]. To experimentally verify the predictions in *Table 1*, the Al$_2$FeCuNiMn HEA was cast, and characterized. The results are shown in *Figure 8*.

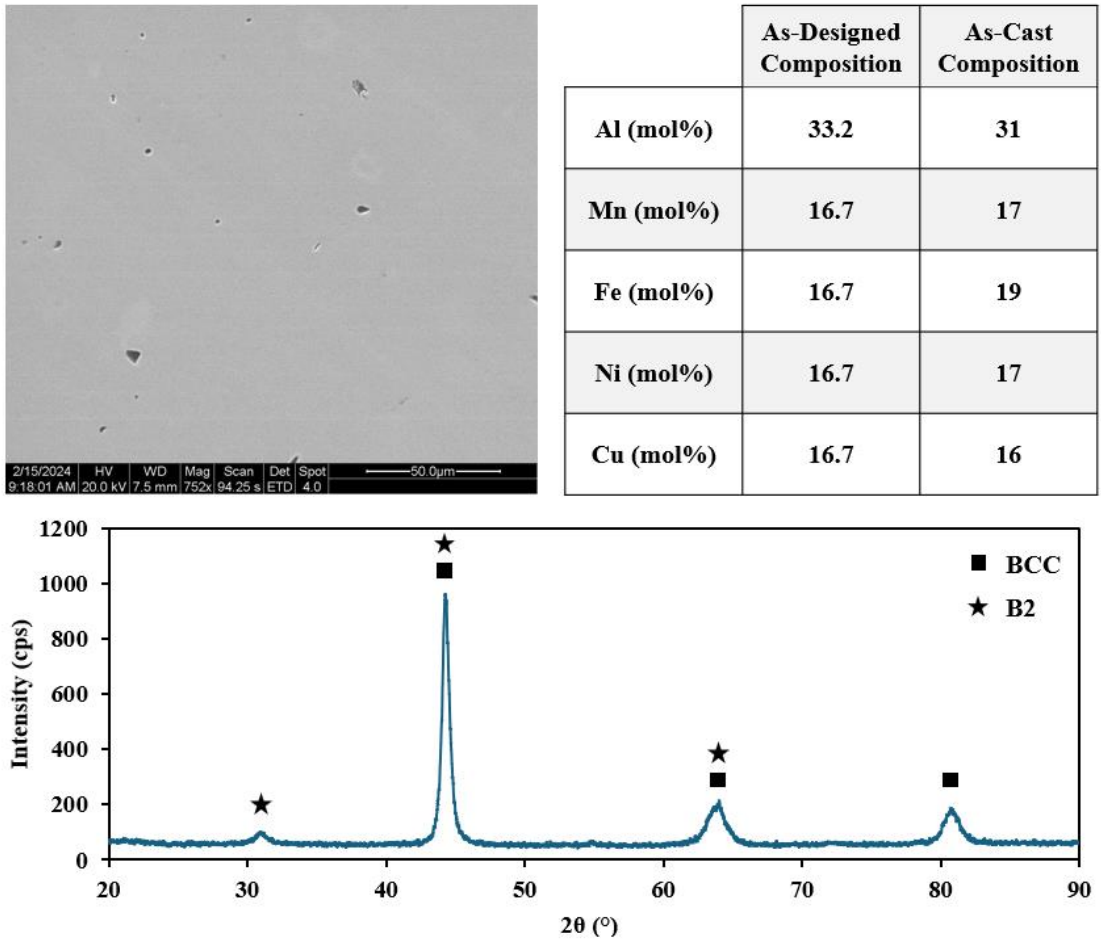

*Figure 8*. Experimental characterizations of as-cast Al$_2$FeCuNiMn HEA: (top left) SEM micrograph, (top right) XRF spectroscopy, and (bottom) XRD pattern

To verify the post-cast composition of the alloy, XRF analysis was done on cast samples. As shown in *Figure 8*, the composition change during casting was negligible. The SEM imaging of the as-cast samples revealed a single-phase microstructure with minor and faint indications of a possible secondary phase. To determine the structures of these observed phases, the XRD pattern of the as-cast alloy was analyzed. As shown in *Figure 8*, the alloy primarily consisted of a BCC phase with a minor B2 phase, and no FCC phase was observed. The presence of ordered B2 structure within a BCC matrix in 3d transition HEAs is a common occurrence, especially at high Al ratios, and has been discussed extensively in the literature [52-56]. These results revealed that the GNN correctly predicted the microstructure of Al$_2$FeCuNiMn HEA (B2 phase was not the method). Despite LSS1 prediction, no FCC phase was detected in the alloy.

## 4. Conclusion

The ALIGNN-FF based approach, proposed in this work, was tested on predicting the structures of 132 different HEAs. The results showed that this approach yields better accuracy as compared to the standard LSS1 prediction based on the VEC values for any given composition of the alloy. The proposed approach was shown to effectively search through the complex potential energy surface configurational space and quantitatively predict the stable crystal structures. To further evaluate the accuracy and applicability of our approach in predicting the structure of an unknown HEA composition, $Al_2FeCuNiMn$ HEA was selected, experimentally cast, and characterized. While the LSS1 method predicted xo-existing of FCC and BCC structures, the ALIGNN-FF based approach predicted the alloy to be BCC, which was verified by the experimental results.

**Conflict of Interest:** The authors state no conflict of interest. This research did not receive any specific grant from funding agencies in the public, commercial, or not-for-profit sectors.

**Acknowledgments:** The authors would like to thank the Materials Engineering Department of California Polytechnic State University, Dr. Trevor Harding, and Mr. Jim Beaver for supporting this research.